\documentstyle[aps,preprint]{revtex}
\tighten	
\begin{document}
\preprint{QMW-AU-97006}
\draft
\title{On the Antenna Pattern of an Orbiting Interferometer} 
\author{{Giacomo Giampieri}\footnote{Present address: Queen
Mary and Westfield College, Astronomy Unit, London E1 4NS.}}
\address{Jet Propulsion Laboratory, California Institute of
Technology, Pasadena, California, 91109}
\date{March 1997}
\maketitle
\widetext
\begin{abstract}

The response of an interferometer changing its orientation
with respect to a fixed reference frame is analyzed in terms of 
the beam-pattern factors and the  polarization-averaged antenna
power pattern. Given the antenna's motion, the latter quantity
describes the antenna's directionality as a function of time.

An interesting case is represented by the class of motions where the
detector's plane is constrained to move on the surface of a cone of
constant aperture; at the same time, the two arms are rotating around
a vertical axis. This picture describes, in particular, the motion of
LISA, a proposed space-based laser interferometer, as well as of other
planned missions. The overall sky's coverage, and that of the galactic
plane in particular, is provided as a function of the cone's aperture.

Similarly, one can consider the case of an earth-based
interferometer. Using the same formalism, one can derive a simple
expression for the antenna pattern, averaged over the time of arrival
of the signal, as a function of the position and orientation on the
earth's surface. In particular, there turn out to be two particular
values for the terrestrial latitude and the inclination angle with
respect to the local parallel which render the time-averaged antenna
response perfectly isotropic.

In the frequency domain, the general result is that the detector's
motion introduces in the instrumental response to a long-duration
continuous signal a few harmonics of the orbital frequency, whose
magnitude depends on the source's position in the sky.  In particular,
we describe LISA's response to circularly polarized sinusoidal waves
coming from a few known binary systems in our Galaxy.

\end{abstract} 

\narrowtext

\section{Introduction}

Gravitational waves  in the low-frequency regime ($10^{-4}$ to
$10^{-1}$ Hz) can only be observed from space, due to terrestrial
disturbances. In space, the only technique currently available,
besides pulsar timing, is based on Doppler tracking of an
interplanetary spacecraft \cite{EW}. However, this relatively
inexpensive method has not provided enough sensitivity, thus far, for
a detection. While better sensitivities may be expected in the near
future, with advanced spacecraft such as CASSINI, much more ambitious
projects for gravitational wave observatories  in space have been
proposed. Among these, the most promising detectors are  based on
space-born laser interferometry. In particular, LISA (Laser
Interferometer Space Antenna) \cite{LISA,Hough}, and OMEGA (Orbiting
Medium Explorer for Gravitational Astrophysics) \cite{OMEGA} consist
of six drag-free, laser-bearing spacecraft, launched in orbit around
the sun (LISA) or the earth (OMEGA). The six spacecraft would be
placed, in pair, at the vertex of a triangle with $5\times 10^6$ km
sides for LISA, and 5 times smaller for OMEGA. At each corner,
the two spacecraft are phase locked through the exchange of a laser
signal, replacing in this way the central mirror of an ordinary 
Michelson intereferometer. Each of the two probes sends a laser beam
to a probe at each of the other two equilateral points, where the
tracking signal is transponded back by phase-locked lasers, and the
returning beams are eventually interfered.

In order to keep the triangular constellation as stable as possible, 
elaborated orbits have been designed. In LISA case each spacecraft is
orbiting  a circle of radius $3\times 10^6$ km over a period of 1 yr.
The interferometer plane, at an inclination of $60^o$ with respect to
the ecliptic, is also rotating around the sun with the same
periodicity. In OMEGA case the orbital plane is almost coincident with
the ecliptic, and the interferometer is rotating around itself with a
period of 53.21 days.

The complicate motion is reflected in the time evolution of the
interferometer's response to a source located in a fixed position in
the sky.  We will investigate the behavior of the antenna response in
presence of a generic motion, and apply our results to the specific
motions of interest. As a side-product of our analysis, we can also
examine a terrestrial interferometer, where the motion is simply
related to the earth's rotation around its axis, and study its antenna
pattern as a function of the location and orientation on the earth's
surface.

In this section, we briefly recall the formalism describing the
antenna response to a gravitational wave passing by, in the
long-wavelength approximation \cite{T300}.

First, we introduce the wave symmetric trace-free (STF) tensor
\begin{equation}
\bbox{W} = h_+ \Re (\vec{m} \otimes \vec{m}) + h_\times \Im
(\vec{m} \otimes \vec{m})\,,
\end{equation}
where the (complex) vector $\vec{m}$ is defined in terms of the
polarization vectors $\vec{e}_X$ and $\vec{e}_Y$ as
\begin{equation}
\vec{m} = \frac{1}{\sqrt{2}} \left( \vec{e}_X + i \vec{e}_Y \right)\,.
\end{equation}
The tensor $\bbox{W}$ represents the wave field as measured in the
interferometer's proper rest frame.
Then we define the STF detector tensor
\begin{equation}
\bbox{D}= \vec{n}_1 \otimes \vec{n}_1 - \vec{n}_2 \otimes
\vec{n}_2\,, 
\end{equation}
where $\vec{n}_i$ is the unit vector along the i-th arm.
The interferometer response is the scalar obtained from the
contraction of the wave tensor $\bbox{W}$ with the detector
tensor $\bbox{D}$
\begin{equation}
R(t)= W_{ij} D^{ij} \equiv F_+ h_+ + F_\times h_\times\,.
\label{response}
\end{equation}
The beam-pattern factors $F_+$ and $F_\times$ depend on the
antenna's orientation with respect to the wave's propagation
direction and polarization axes.

We can choose the reference frame as in
Fig.\ 1, with the x-axis of the $(x,y,z)$ frame bisecting the
interferometer's arms, so that the only non-null components of
$\bbox{D}$ in this reference frame are  
\begin{equation}
D_{12} = D_{21}= \sin(2\Omega)\,,
\end{equation}
where $2\Omega$ is the aperture angle. Therefore, 
in order to obtain $R(t)$, we just need the component
$W_{12}$ in this particular frame. Fig.\ 1
also shows the Euler's angles $\theta,\phi$, and $\psi$ which
transform from the interferometer's frame $(x,y,z)$ to the wave
reference frame $(X,Y,Z)$. The latter  is defined with the Z-axis
opposite to the propagation direction, and the $X$ and $Y$ axes along 
$\vec{e}_X$ and $\vec{e}_Y$, respectively. 

It is easy to find, for the + polarization
\begin{eqnarray}
F_+ &=& \sin(2\Omega)  \Bigl[ \cos\theta\cos(2\phi) \sin(2\psi)
\Bigr.\nonumber\\ &\mbox{}& \Bigl.
\mbox{\hspace{1truecm}} 
+ \frac{1}{2} \left(1 + \cos^2\theta\right) \sin(2\phi)
\cos(2\psi)\Bigr]\,.
\label{F+}
\end{eqnarray}
The beam-pattern factor $F_\times$ is obtained from Eq.\ (\ref{F+})
with the substitution $\psi \rightarrow \psi+\pi/4$, a well known
polarization property of gravitational waves. However, when  averages
over the polarization angle $\psi$ are considered, we can assume,
without loss of generality, $h^+=h^\times=h$. The quantity of interest
is thus the polarization-averaged antenna's {\it power pattern} 
\begin{equation}
P\equiv \langle\left(R\over \sin(2\Omega) h\right)^2 \rangle_\psi\,,
\label{power}
\end{equation}
which for the interferometer in Fig.\ 1 reads 
\begin{equation}
P(\theta,\phi) = \frac{1}{8}\left[ 1 + \cos^4\theta + 6\cos^2\theta -
\sin^4\theta \cos(4\phi)\right]\,.
\label{fixed}
\end{equation}
For future reference, note that this definition of the antenna pattern
is not normalized to unity, the average  of $P(\theta,\phi)$ over the
whole sky being 2/5. A plot of $P(\theta,\phi)$ in polar coordinates is
shown in Fig.\ 5a.

Eq.\ (\ref{fixed}) gives the instantaneous power pattern for a
wave impinging from the direction $(\theta,\phi)$ in the
interferometer's reference frame. If the detector is moving with
respect to the source, then, apart from Doppler effects
considerations, all we need to do is simply replace $\theta$ and
$\phi$ with the appropriate functions $\theta(t)$ and $\phi(t)$. For
example, if the antenna is rotating around its vertical axis 
with angular velocity $\omega$ (Fig.\ 2), then we can obtain the
antenna pattern at any time $t$ from Eq.\ (\ref{fixed}), with the
substitution 
\begin{equation}
\phi \rightarrow \phi - \xi_0 - \omega t\,,
\label{phi}
\end{equation}
where $\xi_0$ is some initial angle. As a matter of fact,  this very
simple case  describes, with good approximation, the time
evolution of  OMEGA \cite{OMEGA}. As mentioned in the Introduction, the
idea of OMEGA is essentially similar to that of LISA, except that the
six spacecraft are launched into a circular earth orbit, 
beyond the Moon orbit. The triangle has now $10^6$ km sides, and it is
rotating around itself with  a period of $\sim 53$ days. 

However, to mask the motion of the antenna with the apparent
motion of the source is not always convenient,
especially when dealing with a large number of sources, or when the
motion is very complicate. In this paper, we shall introduce a more
useful representation, where every quantity is referred to a fixed
reference frame, so that the source's polar coordinates $\theta$ and
$\phi$ remain constant, and the antenna response depends on time
through the actual motion of the interferometer. 

\section{Antenna Pattern for a generic motion}

We now introduce an arbitrary reference frame $(x',y',z')$, with the
only requirement to be stationary with respect to fixed
stars\footnote{Here and in the following, when we say `stationary'
(or `fixed') we mean stationary over the characteristic time scale of
the detector's motion.}. To be more explicit, when dealing with a
space-born interferometer, we can adopt an Ecliptic coordinate
system.  In the last section we will also consider a terrestrial
interferometer, which is most easily described in an Equatorial frame.

The full description of the antenna response requires six
Euler's angles, defined as in Fig.\ 3. The orthogonal transformation
from the wave's frame $(X,Y,Z)$ to the fixed frame $(x',y',z')$ is given
by the orthogonal matrix 
\widetext
\begin{equation}
\bbox{A}=\left(
\begin{array}{ccc}
 \cos\phi\cos\psi - \cos\theta \sin\phi \sin\psi &
-(\cos\phi\sin\psi + \cos\theta \sin\phi \cos\psi) &
\sin\theta \sin\phi \\
 \sin\phi\cos\psi + \cos\theta \cos\phi \sin\psi &
-\sin\phi\sin\psi + \cos\theta \cos\phi \cos\psi &
-\sin\theta \cos\phi \\
 \sin\theta\sin\psi & \sin\theta \cos\psi & \cos\theta
\end{array}
\right)\,.
\label{matrix}
\end{equation}
\narrowtext
The  matrix  $\bbox{B}$ which transform from $(x,y,z)$ to
$(x',y',z')$ is analogous to the matrix $\bbox{A}$, 
with  the Euler's angles $\theta,\phi,\psi$ replaced by  the
corresponding ones $\zeta,\eta,\xi$. Thus, the complete
transformation from the wave's frame to the detector's one is given
by $\bbox{B}^T\cdot \bbox{A}$. Actually, as we may expect from Fig.\ 3,
the angles $\phi$ and  $\eta$ appear in our results only in the
combination $\delta \equiv \phi - \eta$.

After a rather lengthy calculation, one ends up with the following
expressions for the beam-pattern factors $F_+$ and $F_\times$
\begin{mathletters}
\label{F}
\begin{eqnarray}
F_+ &=& \sin(2\Omega) \left[ A \cos(2\xi) \cos(2\psi) + B
\cos(2\xi) \sin(2\psi) +\right.\nonumber\\
 &\mbox{} & \left. 
\mbox{\hspace{1truecm}}
+ C \sin(2\xi)
\cos(2\psi) + D \sin(2\xi) \sin(2\psi) \right] \,,
\\
F_\times &=& \sin(2\Omega) \left[ B \cos(2\xi) \cos(2\psi) - A
\cos(2\xi) \sin(2\psi) +\right.\nonumber\\
&\mbox{} & \left. 
\mbox{\hspace{1truecm}} 
+ D \sin(2\xi) \cos(2\psi) -C \sin(2\xi) \sin(2\psi) \right] \,.
\end{eqnarray}
\end{mathletters}
The coefficients $A,B,C,$ and $D$ in Eqs.\ (\ref{F}) depend only on the
angles $\zeta, \theta,$ and $\delta$. They are explicitly given in
Appendix A.

We recall that the power pattern $P$ is obtained squaring, and
averaging over the polarization angle $\psi$, the interferometer
response. From Eqs.\ (\ref{response}), (\ref{power}), and
(\ref{F}) we eventually obtain 
\begin{equation}
P= \sum\limits_{n=0}^4 \left\{ \left[ \lambda_n + \mu_n \cos(4\xi)
\right] \cos(n\delta) + \sigma_n \sin(4\xi) \sin( n \delta)\right\}
\,,
\label{sum}
\end{equation}
where now the coefficients $\lambda_n, \mu_n,$ and $\sigma_n$
depend only on $\theta$  and $\zeta$. These coefficients, given in
Appendix B, are quite complicate trigonometric
polynomials of their arguments. Nonetheless,
Eq.\ (\ref{sum}) turns out to be very useful in practice. In fact, for the
planned detectors considered in the present paper, the angle $\zeta$
is constant, and thus the only possible time evolution is related to  the
sinusoidal functions of $\xi$ and $\delta$ which appear explicitly in
Eq.\ (\ref{sum}). Before analyzing in more detail the proposed
space-born interferometers, we consider a trivial application of
Eq.\ (\ref{sum}).

\subsection*{A simple example}

As a first test of Eq.\ (\ref{sum}) we can consider, as we did at the
end of Sec.\ I, an interferometer which is rotating around its z-axis 
(see Fig.\ 2), so that its trivial motion is described by 
\begin{equation}
\zeta=\eta=0, \qquad\xi=\xi_0+\omega t\,.
\label{simple}
\end{equation}
{}From Eq.\ (\ref{simple}),
and Eqs.\ (B.1)-(B.15) of Appendix B, we find that the only
non-null coefficients in Eq.\ (\ref{sum}) are
\begin{eqnarray}
\lambda_0 &=& \frac{1}{8}\left( 1 + \cos^4\theta +
6\cos^2\theta\right) \,,
\\ 
\mu_4 &=& \sigma_4 = -\frac{1}{8} \sin^4\theta \,,
\end{eqnarray}
so that
\begin{equation}
P = \frac{1}{8} \left[ 1 + \cos^4\theta + 6 \cos^2\theta -
\sin^4\theta \cos\left[4(\phi-\xi_0-\omega t)\right]\right] \,.
\label{rot}
\end{equation}
As pointed out in Sec.\ I, this result can be 
obtained much more easily directly from Eq.\ (\ref{fixed}), with the
substitution (\ref{phi}).  Eq.\ (\ref{rot}) gives, with good
approximation, the antenna pattern of OMEGA, with
$\omega\simeq  1.4\times 10^{-6}$ sec$^{-1}$.

\section{Space interferometers: Conical motion}

In the previous section, we have considered the antenna power
pattern associated to an unspecified motion of the detector. We will
now focus our attention to the case of a space interferometer, which
presents,  independently of the particular project under
investigation, some very general and interesting properties.

Inserting the space interferometer in
its orbit and keeping the interferometer configuration stable over the
mission lifetime - at least two orbital periods - is a very demanding
navigation task, due principally to the perturbations from the earth and
the other planets. For instance, one of the consequences of the
complicate orbit is the fact that we cannot maintain equal distances
between the probes.   In a recent paper \cite{GG}, we modeled the
noise that is introduced into the differenced data because of the
unequal arms, and showed that the final accuracy of the
interferometer is not compromised. Another example of the problems
we may face in a space-born interferometer is that, due to the earth
disturbances,  high Doppler rates would result. Hellings et al.\
\cite{RH} described a laser  transmitter and receiver hardware
system that provides the required readout accuracy and implements
a self-correction procedure for the on-board frequency standard
used for laser phase measurement.

We will now consider the interferometer's orbit, and
discuss its implication on the antenna response to a wave coming
from a given direction in the sky. In particular, as we mentioned in
the Introduction, LISA \cite{LISA} will orbit the sun at the
earth's distance, as far behind the earth as possible. The  plane
containing the six probes, during its orbit, will remain always tangent
to the surface of a cone of $60^o$ aperture, and  the detector itself
will rotate in this plane with same periodicity - one year - but opposite
direction. Fig.\ 4, reproduced from \cite{Hough}, shows
LISA configuration.

In this section, we will consider a LISA-like motion, characterized by
a generic cone aperture. In other words,  the motion of each of the
three interferometers is assumed to be  described by\footnote{we
could eliminate one of the two initial conditions $\xi_0$ or $\eta_0$
by simply rescaling the time, taking for instance the origin of time at
the passage through the line of nodes ($\xi_0 =0$) or through the
vernal equinox ($\eta_0=0$).} 
\begin{mathletters}
\label{lisa}
\begin{eqnarray} 
\zeta&=& \text{const} \,, 
\\
\eta &=& \eta_0 + \omega t \qquad (\Rightarrow \delta=
\phi-\eta_0 - \omega t) \,,
\\ 
\xi &=& \xi_0 - \omega t \,.
\end{eqnarray}
\end{mathletters}
Note that $\eta$ and $\xi$ are counter-rotating. 
Note also that the three interferometers formed by the triangular
configuration have initial values $\xi_0$ which differ  from each
other by $120^o$, whereas $\eta_0$ is the same for all of them. 
For the sake of conciseness, since
we are concerned here with the antenna power pattern, we will consider
only one of the interferometers, leaving the possibility of exploiting 
the polarization sensitivity to future works.

The proposed LISA orbit has $\zeta=60^o$, a critical value for
the stability of the triangular configuration. Since the behavior of the
antenna, in terms of sky's coverage, directionality, etc., is very
sensitive to the inclination, we will keep $\zeta$ as a free parameter
throughout this paper, and refer to Eqs.\ (\ref{lisa}) as describing 
a `conical' motion.
For a given $\zeta$, the coefficients $\lambda_n, \mu_n,$ and
$\sigma_n$ are now functions of $\theta$. They are explicitly given
in Appendix C for the LISA case.

In order to determine the sky's coverage  during the detector's
lifetime, we need to consider the time average of the antenna
pattern $P$ over one orbital period, which gives
\begin{equation}
\langle P \rangle_T = \lambda_0 + \frac{1}{2}\left(
\mu_4+\sigma_4\right) \cos[4(\phi-\eta_0-\xi_0)] 
\label{time}
\end{equation}
Fig.\ 5 shows a plot of $\langle P \rangle_T$ in polar coordinates for
various values of the cone's aperture $\zeta$.  Note that $\zeta=0$
corresponds to an interferometer fixed in space; since we are taking
$\xi$ and $\eta$ counter-rotating, the interferometer does not rotate
at all.

We know from Eq.\ (\ref{fixed}) that an interferometer fixed in
space can never detect waves impinging from four specific
null-directions, given by $\theta=\pi/2$, and $\phi=k \pi/2\; (
k=0,\dots,3)$. We can now analyze what happens to these
null-directions in the generic conical case, focussing our attention to
the ecliptic plane $\theta=\pi/2$. It is easy to show, from the
behavior of the functions $\lambda_0$ and $|\mu_4+\sigma_4|/2$ (see
Fig.\ 6), that there is a  tendency for
the null-directions on the ecliptic plane to remain visible, although
the magnitude of this effect is strongly affected by the value of
$\zeta$: for $\zeta=0$ the null-directions are obviously completely
preserved, while for $\zeta=\pi/2$ the $\phi$-dependence is very
poor. LISA is much closer to the latter case, and actually its $\langle
P\rangle_T$ is almost independent on $\theta$ as well, as we shall
see in a while.

Note that, if $\xi$ and $\eta$ were
co-rotating, instead of counter-rotating as in LISA, then in the last
term of Eq.\ (\ref{time}) we would have to make the substitution  
\begin{equation}
\frac{1}{2}\left(\mu_4+\sigma_4\right)
\rightarrow 
\frac{1}{2}\left(\mu_4-\sigma_4\right) \,.
\end{equation}
which means that the $\phi$-term would be even smaller, in
absolute value, compared to $\lambda_0$.

One might infer from Fig.\ 5d that, if the inclination is close to
$\zeta=\pi/2$, then the zeros of the antenna pattern can be found in
the direction orthogonal to the ecliptic plane. However, this is not the
case, since $\theta=0$ implies that $P$ is still given by 
Eq.\ (\ref{fixed}),
with $\theta$ and $\phi$ replaced, respectively, by $\zeta$ and
$\xi$. Thus, even for $\zeta=\pi/2$ the time average is
non-zero at the poles.

Another important issue related to Eq.\ (\ref{time}) is the
time-averaged antenna directionality. As we can predict from Fig.\ 5,
directionality is strongly dependent on the angle $\zeta$. To make
this statement more precise, let us consider the r.m.s.\ deviation
from isotropy, defined as
\begin{equation} 
\Delta \equiv \left[ \frac{175}{48\pi}
\int\limits_{4\pi} \left(\langle P\rangle_T - 2/5\right)^2 d\Omega
\right]^{1/2} \,.
\label{delta}
\end{equation}
The normalization factor in front of Eq.\ (\ref{delta}) is chosen in such a
way that $\Delta$ is normalized to one for a fixed interferometer.
For a conical motion, inserting Eq.\ (\ref{time}) 
in Eq.\ (\ref{delta}) we easily obtain
$\Delta=\Delta(\zeta)$, shown in Fig.\ 7. Explicitly, we find
\begin{eqnarray}
\Delta(\zeta) &=& \frac{1}{288} \left[ 19779 + 120 \cos\zeta -
118380 \cos^2\zeta 
\right.\nonumber\\
&\mbox{}&\left.
+ 840 \cos^3\zeta + 180690 \cos^4\zeta
+840 \cos^5\zeta 
\right.\nonumber\\
&\mbox{}&\left.
-3180 \cos^6\zeta +120 \cos^7\zeta +2115
\cos^8\zeta\right]^{1/2}\,.
\end{eqnarray}

Thus, the antenna pattern is distributed more and more isotropically
as we  increase $\zeta$ from $\zeta=0$. After we
reach a minimum at $\zeta\simeq 55^o$, the antenna's directionality
starts increasing again. 

Therefore, we can conclude that, in the conical case,  it is impossible to
get a perfectly isotropic response, i.e.\ we never get $\Delta=0$ (see
also the above discussion about the $\phi$-dependence). However, we
can get very close to this ideal situation, if we choose
$\zeta$ appropriately. Remarkably, LISA's inclination is very close to
the optimal value $\zeta\simeq 55^o$, and gives $\Delta(60^o)
\simeq 0.14$. This means that LISA, during its lifetime, will cover the
whole sky in an approximately uniform manner. Of course, in some
circumstances, directionality needs to be preserved. For instance, one
may want to disentangle the isotropic component of the stochastic
background from the anisotropic contribution of the galactic binaries.
Directionality can always be  preserved integrating over a shorter
period. Fig.\ 8 shows LISA's antenna pattern averaged over 3 and 6 
months, respectively.

Finally, we want to consider the  interferometer's responsiveness to
the galactic plane, given that most of the strongest sources will lie on
this plane. In our coordinate system, the galactic plane is
characterized by 
\begin{equation}
\theta(\phi) = \arctan(\alpha \cos\phi + \beta\sin\phi)^{-1}\,,
\label{plane}
\end{equation}
where $\alpha \simeq 1.75$ and $\beta \simeq 4\times10^{-4}$.
Assuming, for simplicity, that the sources are distributed
isotropically on the plane, the event rate of disk's sources is
proportional to the average area of the intersection of this plane
with  the  antenna pattern, equal to 
\begin{equation} 
G = \frac{1}{2T} \int\limits_0^{2\pi} \int\limits_0^T
 \Bigl[ P(\theta(\phi),\phi) \Bigr]^2 dt\, d\phi \,,
\end{equation}
where $\theta(\phi)$ is the function given in Eq.\ (\ref{plane}). This area
depends on the details of the detector's motion, in our case on
$\zeta$,  $\xi_0$, and $\eta_0$. Fig.\ 9 shows the
quantity $G$ as a function of $\zeta$, since the dependence on
$\xi_0$ and $\eta_0$ can be neglected in a first approximation.
For LISA, we find  the small value $G \sim 0.46$, the exact
value  depending on the initial conditions $\eta_0+\xi_0$. We
conclude that LISA is not particularly sensititive to the galactic disk,
due to the fairly large inclination to the Ecliptic of both the detector's
plane and the Galaxy. According to this crude analysis, we can expect
that OMEGA, with a smaller, almost negligible, inclination, would
increase its chances of observing a signal from the disk by roughly a
factor two. Note, however, that previous calculations 
\cite{Evans,Hils} have shown that
gravitational radiation from galactic binaries in the disk is
comparable to that coming from the local region ($r < 200$ pc). Since
the latter is isotropically distributed, the time-varying signal from
the disk contributes only a fraction of the galactic binaries
stochastic background. Nonetheless, this small component could make
the stochastic signal distinguishable from the detector's noise. In
a future paper we will investigate the
amplitude modulations introduced by the antenna motion in the 
confusion noise generated by different populations of galactic binary
systems, and describe how to exploit this effect in order to detect the
signal and to obtain information about the distribution of
sources in the Galaxy.

\section{Earth-based interferometers}

As an additional application of Eq.\ (\ref{sum}), let us consider a 
terrestrial interferometer. 
In this case the most convenient choice for the `fixed'
reference frame is  the Equatorial one, with the $z'$ axis directed
toward the North Pole, and the $x'$ axis toward the Vernal Equinox.
In this frame, the motion of the interferometer becomes similar to
the conical case previously analyzed, except that now the detector can
not rotate on itself, of course. In other words we have  
\begin{mathletters}
\label{earthmotion}
\begin{eqnarray}
\zeta &=& \pi/2 -\ell \,,
\\
\eta &=& \eta_0 +\omega t \,, 
\\
\xi &=& \iota \,,
\end{eqnarray}
\end{mathletters}
where  $\ell$ is the terrestrial latitude,  $\omega$ is the earth's
angular velocity of rotation, and 
$\iota\in[-\frac{\pi}{4},+\frac{\pi}{4}]$ is the angle between the
arms' bisector and the local parallel.

For example, let us consider
the average of the antenna pattern $P$  over the time of arrival of 
the signal. In the terrestrial case, as opposed to the conical case
considered in Sec.\ III, the interferometer cannot rotate around its
vertical axis, and therefore averaging over time or $\phi$ gives the
same result, namely
\begin{equation} 
\langle P \rangle_T = \lambda_0
(\theta,\ell) + \mu_0(\theta,\ell) \cos(4\iota)\,.
\label{earth}
\end{equation}
For any specific value of $\theta$, for example  $\theta=102^o$,
corresponding to the direction of the center of the Virgo cluster, 
Eq.\ (\ref{earth}) gives
the square of the r.m.s.\ power as a function of the antenna's position
and orientation on the earth's surface, a quantity already numerically
studied in \cite{ST}. 

Instead of fixing $\theta$, we could  try to answer the
question: is there any particular location and orientation
for which  the antenna pattern, averaged over one day, is
isotropic? The answer turns out to
be affirmative, and the isotropic antenna is characterized by 
\begin{mathletters}
\label{isotropy}
\begin{eqnarray}
\ell_{is} &=& \arcsin \left(\pm \frac{1}{\sqrt{3}}\right) 
\simeq \pm 35^o.26438972 \,,\\
\iota_{is} &=& \frac{1}{4}\arccos \left(-\frac{1}{5}\right) 
\simeq \pm 25^o.38423976 \,.
\end{eqnarray}
\end{mathletters}
One can easily check, by inspection, that $\ell_{is}$ and
$\iota_{is}$ produce $\langle P \rangle_T \equiv 2/5$ or,
equivalently, $\Delta\equiv 0$. A detector located at latitude
$\ell_{is}$ and oriented by $\iota_{is}$ maximizes the event rate of
an isotropic population of sources.

In the event that the detector's position has already been chosen, one
can still make use of Eq.\ (\ref{earth}) in order to find the optimal
orientation $\iota_*$ which gives the least directional antenna pattern
at that latitude. 
At each latitude $\ell$,  we define as optimal that orientation
$\iota_*$ which minimize the quantity $\Delta$.
Fig.\ 10 shows $\iota_*=\iota_*(\ell)$ and the corresponding
minimum $\Delta_*\equiv \Delta(\ell,\iota_*)$. 

We stress that, as one may actually expect, $\Delta$ depends
much more strongly on $\ell$ than $\iota$, and in particular the
antenna becomes rapidly anisotropic as we move away from
$\ell_{is}$, no matter how optimally we try to choose $\iota$.
Moreover, we can foresee several terrestrial interferometers to be
operative in the near future, so that the sky's coverage of a single
antenna is not really an issue as critical as in the space-born case
previously discussed.

\section{Sinuosoidal waves from binaries. Fourier analysis.} 

It is generally assumed that galactic and extragalactic binary systems
are the most promising sources of gravitational waves for detectors
based on laser interferometry. In fact, waves from a binary star,
including the effect of eccentricity, orbital inclination, and also
post-Newtonian corrections, have long been studied, and are today well
understood. In particular, the sensitivity of the planned space
interferometers should allow the detection of waves from several known
galactic binary stars. In the LISA and OMEGA 
frequency band, the strongest among these sources are presumably 
the Interacting White Dwarfs Binaries (IWDB)
\cite{Evans,Hils}. Table I contains the available data for
five IWDB, including the amplitude and frequency of the expected
gravitational waves. We have applied our results to these objects,
and describe the LISA's response to the waves originating from them.
Fig.\ 11 shows the beam-pattern factors $F_+$ and $F_\times$
for the five IWDB in Table I, as seen from LISA over one
year. Fig.\ 11 also shows the analogous quantities for a sinusoidal
signal, of unspecified amplitude and frequency,  coming from the
galactic centre.

We will now consider the effect of the motion in the frequency
domain, for both  the space-born and the terrestrial cases. We define
the Fourier series as usual
\begin{equation}
P(t) = a_0 + \sum\limits_{k=1}^{\infty} \left\{ a_k
\cos\left(k\pi t\over\ L\right) + b_k \sin\left(k\pi t\over\ L\right)
\right\}\,,
\end{equation}
where $T=2L$ is the orbital period, and
\begin{eqnarray}
a_0 &=& \frac{1}{2L}  \int\limits_{-L}^{+L} P(t)  dt\,,\\
a_k &=& \frac{1}{L} \int\limits_{-L}^{+L} P(t) 
\cos\left(k\pi t\over\ L\right) dt \,,\\
b_k &=& \frac{1}{L} \int\limits_{-L}^{+L} P(t) 
\sin\left(k\pi t\over\ L\right) dt\,.
\end{eqnarray}
For the conical case analyzed in Sec.\ III, using eqs.(B1)-(B14) of
Appendix B, one finds that the only non-zero Fourier coefficients are
given by  ($\delta_0 \equiv \phi-\eta_0,\: k=1,\dots,8$):   
\begin{eqnarray}
a_0 &=& \lambda_0 +\frac{1}{2}(\mu_4+\sigma_4)
\cos[4(\delta_0-\xi_0)] \,,\\
a_k &=& \mu_0 \cos(4\xi_0) \delta_{k 4} +
\sum\limits_{n=1}^{4}\Bigl\{\lambda_n \cos(n\delta_0) \delta_{nk} 
\Bigr.\nonumber
\\ &\mbox{}& \Bigl. 
+\frac{1}{2} (\mu_n+\sigma_n) 
\cos(n\delta_0-4\xi_0) \delta_{k+n,4} 
\Bigr.\nonumber
\\ &\mbox{}& \Bigl. 
+\frac{1}{2} (\mu_n-\sigma_n)
\cos(n\delta_0+4\xi_0) \delta_{k-n,4}\Bigr\}\,,
\\
b_k &=& \mu_0 \sin(4\xi_0) \delta_{k 4} +
\sum\limits_{n=1}^{4}\Bigl\{\lambda_n \sin(n\delta_0) \delta_{nk} 
\Bigr.\nonumber
\\ &\mbox{}& \Bigl.
-\frac{1}{2} (\mu_n+\sigma_n)
\sin(n\delta_0-4\xi_0) \delta_{k+n,4}  
\Bigr.\nonumber\\
&\mbox{}& \Bigl.
+\frac{1}{2} (\mu_n-\sigma_n)
\sin(n\delta_0+4\xi_0) \delta_{k-n,4}\Bigr\}\,.
\end{eqnarray}

The analogous calculation for the terrestrial case, described by 
Eqs.\ (\ref{earthmotion}),  gives
the following nine coefficients $(k=1,\dots,4)$ 
\begin{eqnarray}
a_0 &=& \lambda_0 + \mu_0 \cos(4\iota)\,,\\
a_k &=& \left[\lambda_k +\mu_k\cos(4\iota)\right] \cos(k\delta_0)
+\sigma_k \sin(4\iota) \sin(k\delta_0)\,,\\
b_k &=& \left[\lambda_k +\mu_k\cos(4\iota)\right] \sin(k\delta_0)
-\sigma_k \sin(4\iota) \cos(k\delta_0)\,.
\end{eqnarray}

In general, given the initial detector's position, these Fourier
coefficients depend on the source's coordinates $\theta$ - through
$\lambda_k, \mu_k, \sigma_k$ - and $\phi$ - through $\delta_0$. If
the source location is known, then one can look for these spectral
lines  as a convincing signature about the gravitational origin of the
signal. 
When the source's coordinates are
unknown, however, one has to deal with the complication arising
from the Doppler effect \cite{Schutz}. The motion of the detector,
besides the amplitude modulation described in the present work, also
introduces a location-dependent phase modulation, in the form
of a Doppler broadening of the sinusoidal signal. In the case of LISA,
the magnitude of this effect, over a  period $T=1$ yr, is
\begin{equation}
\frac{\Delta f}{f}  \simeq 2\times 10^{-4} \,,
\end{equation}
so that, in the spectral region below $10^{-2}$ Hz, we do not
have any hope of finding the aforementioned lines, separated
from each other by only $1/T\simeq 3\times 10^{-8}$ Hz.  
For a terrestrial interferometer, the situation is analogous, only
complicated by the simultaneous effects of the diurnal and annual
motion of the earth, and also by the earth-moon interaction.

In conclusion, for long enough observations, we need
special techniques to compensate for the frequency spread over
several frequency-resolution bins, and eventually to recover the
amplitude modulation described in this paper. Several different
strategies can be adopted to overcome this problem, although none of
them is completely satisfactory, due to the large amount of
computation involved. See \cite{Schutz} for more details.
In any case, the amplitude modulation can be exploited for an
independent  measurement of the location of the source, and, in
addition, to  obtain the polarization of the wave.

\section{Conclusions}

In the present work, we have considered a gravitational wave
interferometer, in motion with respect to fixed stars, and studied
the resulting amplitude modulation of a long-duration continuous 
signal. The general
results are presented in Sec.\ II, where the istantaneous
beam-pattern factors - Eqs.\ (\ref{F}) - and  the
polarization-averaged antenna power pattern  - Eq.\ (\ref{sum}) -
are given as functions of time, for a generic motion.

Next, two particular
cases have been analyzed: 1) the probable orbit of a space-born
interferometer, with particular emphasis on LISA, and 2) the motion
of a ground-based interferometer.

For what concerns LISA, we have shown that its peculiar motion
makes the time-averaged antenna pattern practically isotropic, thus
providing an uniform coverage of the whole sky over the period of
one year. For shorter integration periods directionality is mostly
preserved, and can be exploited where necessary, for example in the
search of a galactic binaries background. However, when we focused
on the galactic disk, we found that the
average antenna response is far from optimal, due to the relative
orientation of the Ecliptic and the galactic plane itself. We stress that
these results are not conclusive, since we have neglected the
anisotropy in the distribution of the sources with respect to the sun,
due to the fact that we are located near the edge of the disk. In this
respect, additional work is needed.

In the terrestrial case, thanks to the probable redundancy of future
gravitational wave observatories, the discussion about a single
antenna's sky's coverage is not so critical. However, we found that
there are particular positions on the earth's surface, given in Eqs.\
(\ref{isotropy}), which render the time-averaged antenna response
perfectly isotropic. For what concerns the galactic plane, since the
latter makes with the Equatorial plane approximately the same angle
it makes with the Ecliptic ($\sim 60^o$), the result is analogous to the
conical case, with the angle $\zeta$ interpreted as $90-\ell$ in Fig.\
9. In other words, the response to the galactic plane increases as we
move the interferometer from the equator toward the poles, with a
minor role played by the orientation angle $\iota$.

\acknowledgments
The author would like to thank P.Bender and R.Hellings for discussions. 
The research described in this paper was performed while the
author held an NRC-NASA Resident Research Associateship at the Jet
Propulsion Laboratory, California Institute of Technology, under a
contract with the National Aeronautics and Space Administration.

\appendix

\section{}

In this Appendix, we give the coefficients $A,B,C,$ and $D$ which
enter in the expressions of $F_+$ and $F_\times$, Eqs.\ (\ref{F}).
They, in turn, can be expressed in terms of intermediate quantities
$\alpha_1, \alpha_2, \beta_1,$ and $\beta_2$ as follows
\begin{eqnarray}
A &=& \alpha_1 \beta_1 + \alpha_2 \beta_2 \\
B &=& \alpha_1 \beta_2 - \alpha_2 \beta_1 \\
C &=& \frac{1}{2} \left[ \beta_1^2 - \beta_2^2 - \alpha_1^2 +
\alpha_2^2\right]\\
D &=& \alpha_1 \alpha_2 + \beta_1 \beta_2 
\end{eqnarray}
where
\begin{eqnarray*}
\alpha_1 &=& \cos\delta\\
\alpha_2 &=& \cos\theta \sin\delta\\
\beta_1 &=& \cos\zeta \sin\delta\\
\beta_2 &=& \cos\zeta \cos\theta \cos\delta + \sin\zeta \sin\theta
\end{eqnarray*}

\section{}
In this Appendix, we give the coefficients $\lambda_n, \mu_n$, and
$\sigma_n\quad (n=0,\dots,4)$, defined in Eq.\ (\ref{sum}), as
functions of the angles $\theta$ and $\zeta$.
\begin{eqnarray}
\lambda_0 &=& \frac{1}{64} \left( 35 + 35 \cos ^4\theta \cos^4\zeta
- 30 \cos^2\theta + 3\cos^4\zeta + 108 \cos^2\zeta
\cos^2\theta  \right. \nonumber\\
&\mbox{}& \left.  - 30 \cos^2\zeta \cos^4\theta - 30 \cos^4\zeta
\cos^2\theta - 30 \cos^2\zeta + 3 \cos^4\theta\right)
\\
\lambda_1 &=& \frac{1}{32} \sin(2\zeta) \sin(2\theta) \left( 15 -3
\cos^2\theta -3 \cos^2\zeta + 7\cos^2\zeta \cos^2\theta \right)
\\
\lambda_2 &=& \frac{1}{16} \left(7 + 7 \cos ^4\theta \cos^4\zeta
- 8 \cos^2\theta + \cos^4\zeta + 16 \cos^2\zeta \cos^2\theta  
- 8\cos^2\zeta \cos^4\theta \right. \nonumber\\
&\mbox{}& \left. - 8 \cos^4\zeta
\cos^2\theta - 8 \cos^2\zeta + \cos^4\theta\right)
\\
\lambda_3 &=& \frac{1}{8} \sin^3\zeta \cos\zeta \sin^3\theta
\cos\theta
\\ 
\lambda_4 &=& \frac{1}{64} \sin^4\zeta \sin^4\theta
\\
\nonumber
\\
\mu_0 &=&  -\frac{1}{64} \sin^4\zeta 
\left( 3 - 30 \cos^2\theta + 35 \cos^4\theta \right)
\\ 
\mu_1 &=& -\frac{1}{16} \sin(2\theta) \cos\zeta  \sin^3\zeta
\left( 3  - 7\cos^2\theta \right)
\\
\mu_2 &=&  \frac{1}{16}\sin^2\zeta \sin^2\theta
\left(1+\cos^2\zeta\right)  \left(1 -7 \cos^2\theta\right) 
\\ 
\mu_3 &=& \frac{1}{16} \sin(2\zeta) \sin^3\theta \cos\theta
\left(3 + \cos^2\zeta\right)
\\
\mu_4 &=& -\frac{1}{64} \sin^4\theta \left( 1 + \cos^4\zeta +
6\cos^2\zeta\right)
\\
\nonumber
\\
\sigma_0 &=& 0
\\
\sigma_1 &=& -\frac{1}{16} \sin(2\theta) \sin^3\zeta \left( 3
-7 \cos^2\theta  \right)
\\
\sigma_2 &=& \frac{1}{8} \cos\zeta \sin^2\zeta \sin^2\theta \left(1 
 - 7\cos^2\theta\right)
\\
\sigma_3 &=& \frac{1}{8} \sin\zeta \cos\theta \sin^3\theta
\left(1+3\cos^2\zeta\right)
\\
\sigma_4 &=& -\frac{1}{16} \sin^4\theta \cos\zeta
\left(1+\cos^2\zeta\right) 
\end{eqnarray}

\section{}
In this Appendix, we provide the coefficients $\lambda_n, \mu_n$, and
$\sigma_n\: (n=0,\dots,4)$ for LISA
\begin{eqnarray}
\lambda_0 &=& \frac{1}{1024} \left(443 -37 \cos^4\theta -78
\cos^2\theta\right)
\\
\lambda_1 &=& \frac{\sqrt{3}}{256} \sin(2\theta)\left(57-5
\cos^2\theta\right) 
\\
\lambda_2 &=& \frac{9}{256} 
\left(9-\cos^4\theta- 8\cos^2\theta\right)
\\
\lambda_3 &=& \frac{3\sqrt{3}}{128} \sin^3\theta\cos\theta
\\
\lambda_4 &=& \frac{9}{1024}  \sin^4\theta
\\
\nonumber
\\
\mu_0 &=& -\frac{9}{1024} \left(3 + 35 \cos^4\theta -30
\cos^2\theta\right)
\\
\mu_1 &=& -\frac{3\sqrt{3}}{256} \sin(2\theta)\left( 3
- 7 \cos^2\theta\right)
\\
\mu_2 &=& \frac{15}{256} \sin^2\theta \left(1 -7 \cos^2\theta
\right)
\\
\mu_3 &=& \frac{13\sqrt{3}}{128} \sin^3\theta \cos\theta
\\
\mu_4 &=& -\frac{41}{1024}  \sin^4\theta
\\
\nonumber
\\
\sigma_0 &=&0
\\
\sigma_1 &=& -\frac{3\sqrt{3}}{128} \sin(2\theta)\left( 3
-7 \cos^2\theta \right)
\\
\sigma_2 &=& \frac{3}{64} \sin^2\theta\left(1 -
7 \cos^2\theta\right)
\\
\sigma_3 &=& \frac{7\sqrt{3}}{64} \sin^3\theta \cos\theta
\\
\sigma_4 &=& - \frac{5}{128}  \sin^4\theta
\end{eqnarray}

\begin{figure}
\caption{The geometry of the interferometer. Note that the usual
spherical polar coordinates of the source's position are $(\theta,
\phi-\pi/2)$.}
\end{figure}

\begin{figure}
\caption{OMEGA: a space-born detector rotating around its vertical
axis. Although the actual OMEGA's motion is better described with the
formalism introduced in Sec.\ III, this simple example can provide a
good approximation to it.}
\end{figure}

\begin{figure}
\caption{The relation between the wave's $(X,Y,Z)$, the detector's
$(x,y,z)$, and the fixed $(x',y',z')$ reference frames.}
\end{figure}

\begin{figure}
\caption{LISA: a space-born interferometer in orbit around the sun.
Reproduced from \protect\cite{Hough}, with permission.}  
\end{figure}

\begin{figure}
\caption{The antenna pattern, averaged over time and polarization,
for various values of the inclination angle $\zeta$. a) $\zeta=0$. b)
$\zeta=30^o$, c) $\zeta=60^o$, d) $\zeta=90^o$. LISA corresponds to
case c).} 
\end{figure}

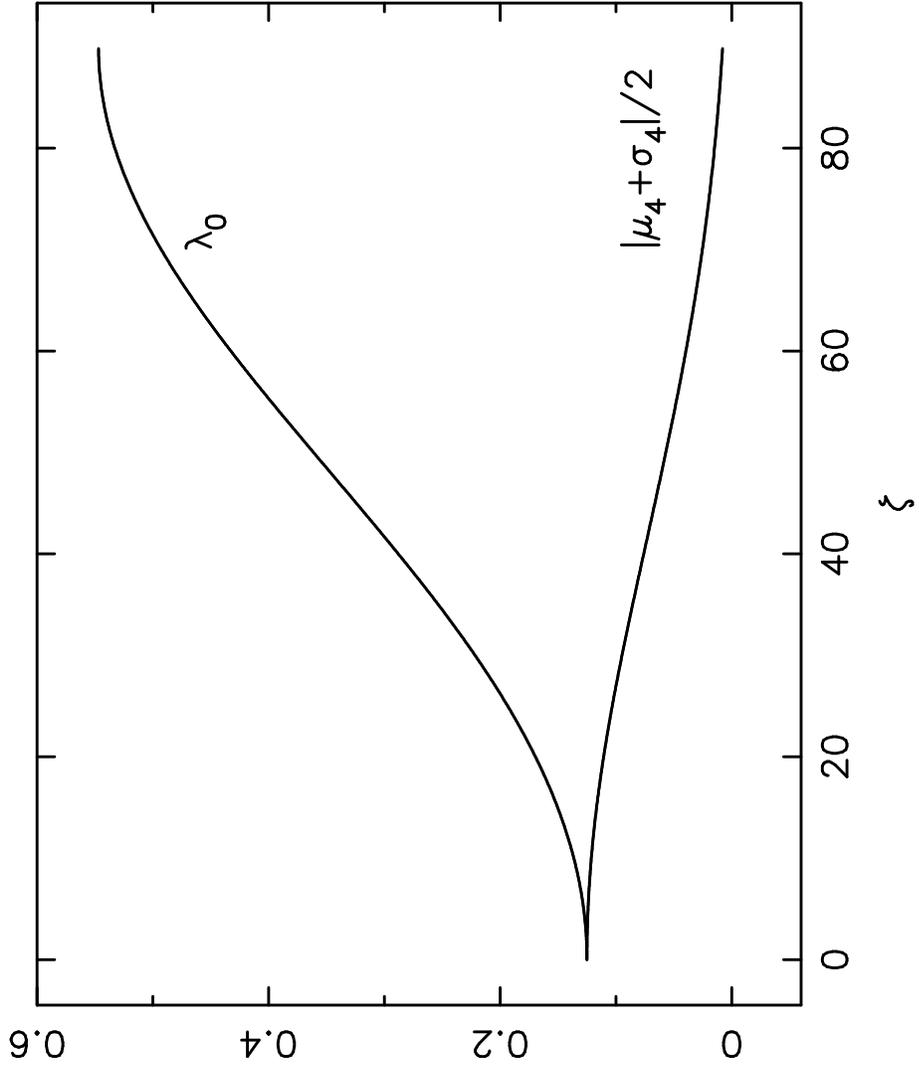
\begin{figure}
\caption{The $\phi$-dependence on the Ecliptic plane, as a function of
the angle $\zeta$. The bigger is
the difference between $\lambda_0$ and $|\mu_4+\sigma_4|/2$, the
smaller is the dependence on the direction $\phi$ in the averaged
antenna pattern.}
\end{figure}

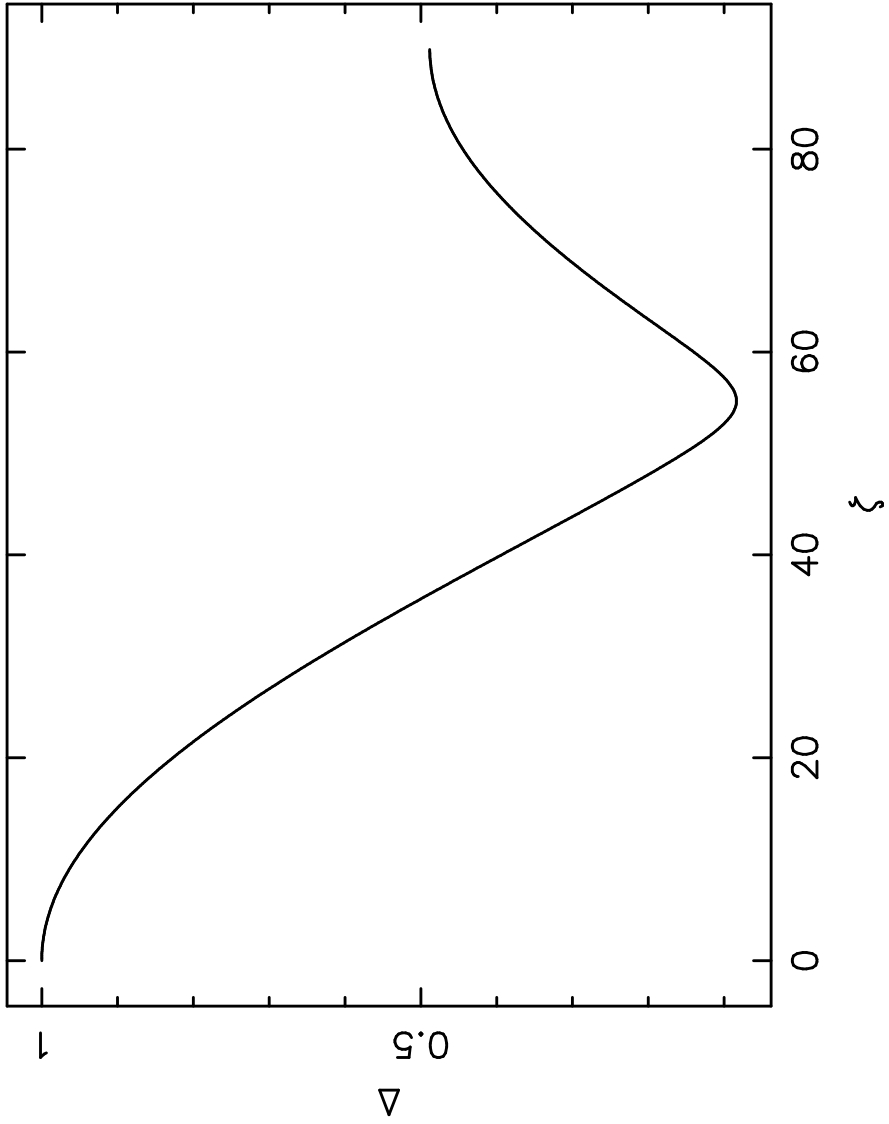
\begin{figure}
\caption{The r.m.s.\ deviation from isotropy of the
time-averaged antenna pattern, as a function of $\zeta$. 
The minimum of $\Delta$ corresponds to the maximum attainable
isotropic response. This minimum occurs at $\zeta\simeq 55^o$, for
which $\Delta\simeq 0.08$. For LISA, $\Delta(60^o)\simeq 0.14$.} 
\end{figure} 

\begin{figure}
\caption{LISA's antenna pattern averaged over (a) 3 months and
(b) 6 months. The initial conditions are $\xi_0=\eta_0=0$.} 
\end{figure}

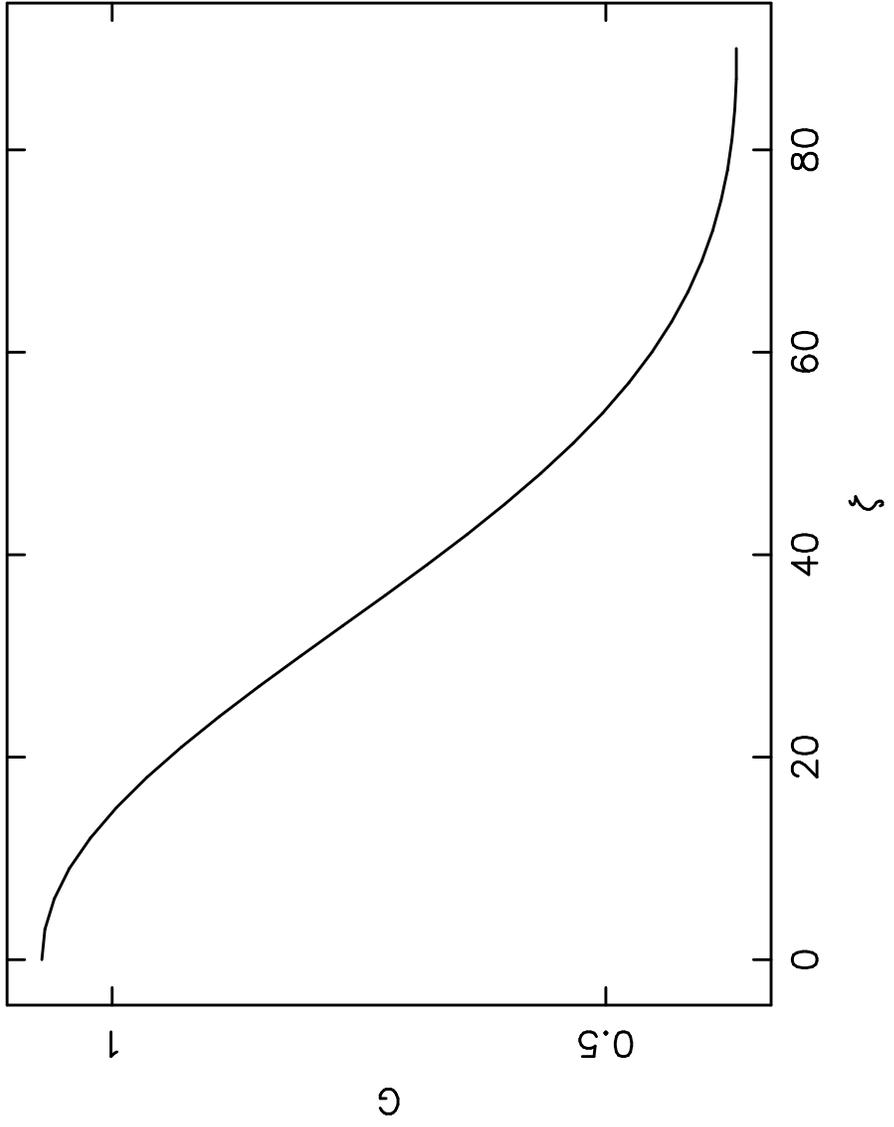
\begin{figure}
\caption{The response of a space interferometer to the galactic disk, 
as a function of the angle $\zeta$.  For simplicity, a small dependence
on the detector's initial orientation, more precisely on
$\eta_0+\xi_0$, has been neglected.}
\end{figure}

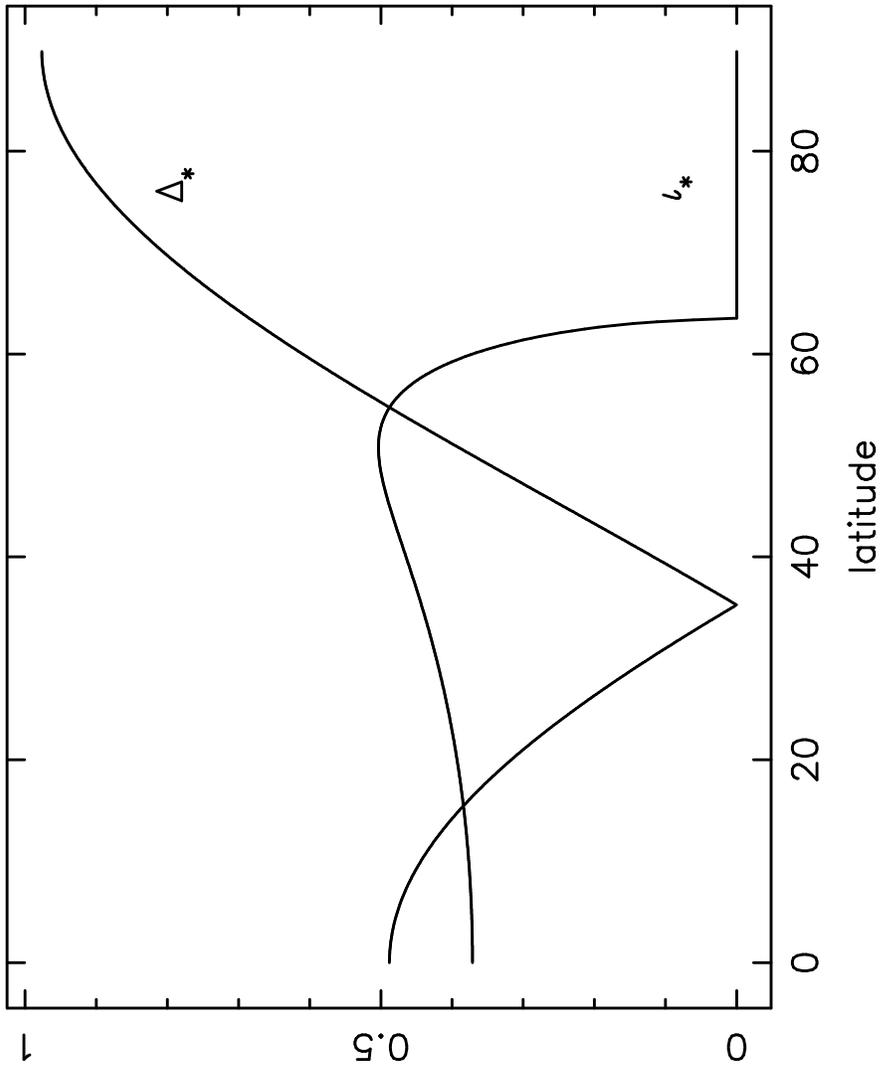
\begin{figure}
\caption{The minimum r.m.s.\ deviation from isotropy $\Delta_*$
attainable at any given latitude. The corresponding optimal
orientation $\iota_*$ is also shown.}
\end{figure}

\begin{figure}
\caption{The LISA normalized responses to the 5 IWDB systems
shown in Table I, and to a hypothetical 
source in the galactic center, over one year of observations. The two
curves give the beam-pattern factors $F_+$ (solid line) and $F_\times$
(dashed line). We have assumed the initial conditions
$\xi_0=\eta_0=0$, and a polarization angle $\psi=2$. Horizontal axis
is time in years.}
\end{figure}

\begin{table}
\caption{Data for 5 known IWDB. The first column is the name, the
second and third column are, respectively, the angles $\theta$ and
$\phi$ in the Ecliptic coordinate system. The last two columns gives,
respectively, the predicted amplitude and frequency of the
gravitational waves.} 
\begin{tabular}{ccccc} 
Name & $\theta$ & $\phi$ & GW Amplitude & GW Frequency \\
 & (degrees) & (degrees) & ($10^{-22}$) & ($10^{-3}$ Hz) 
\\ \tableline
{\it AM CVn} & 52.56 & 260.38 & 5.27 & 1.94 \\ 
{\it CR Boo}    & 72.10 & 292.27 & 2.82 & 1.34 \\  
{\it V803 Cen} & 120.31 & 306.17 & 0.89 & 1.24 \\ 
{\it CP Eri} & 120.83 & 151.77 & 4.02 & 1.16 \\ 
{\it GP Com} & 67.00 & 277.73 & 1.77 & 0.72   \\ 
\end{tabular}
\end{table}
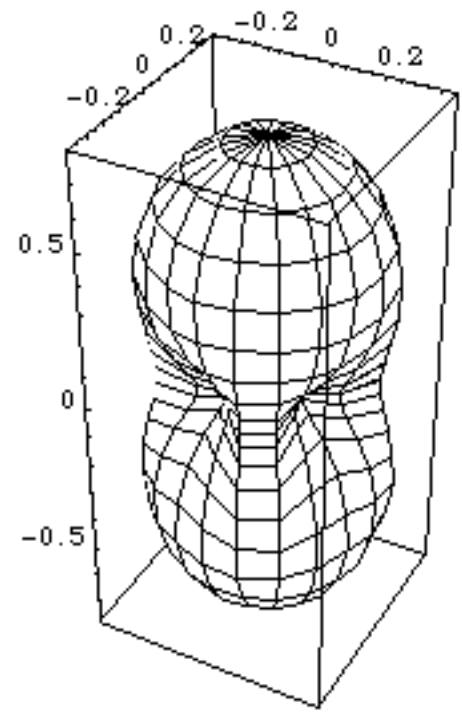
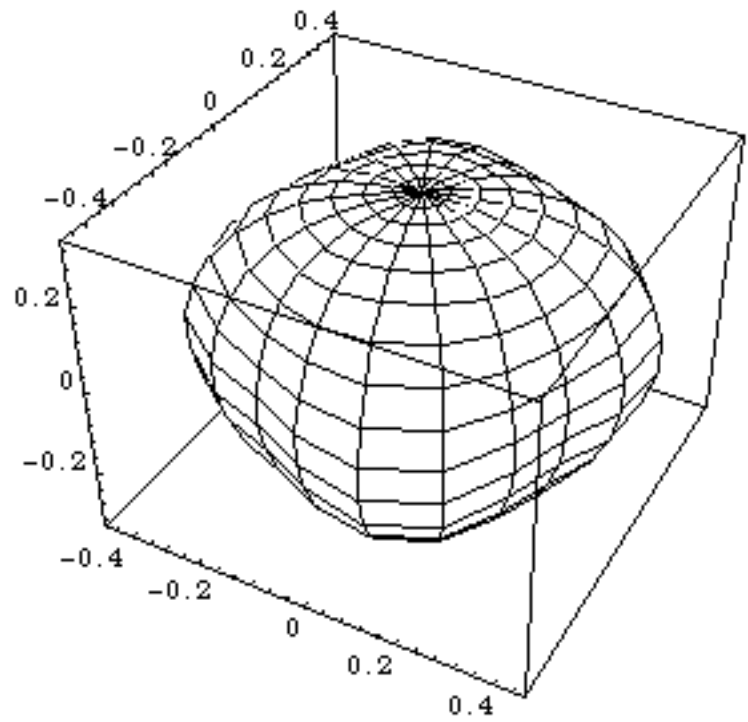
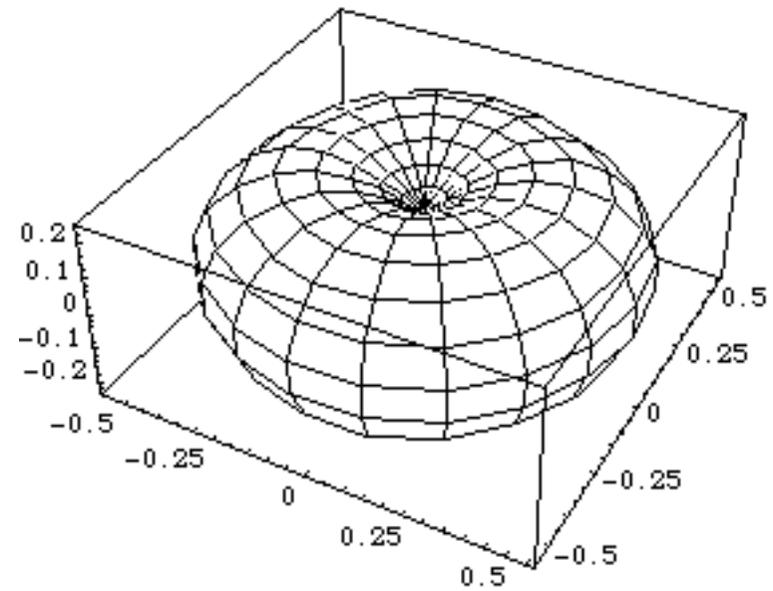
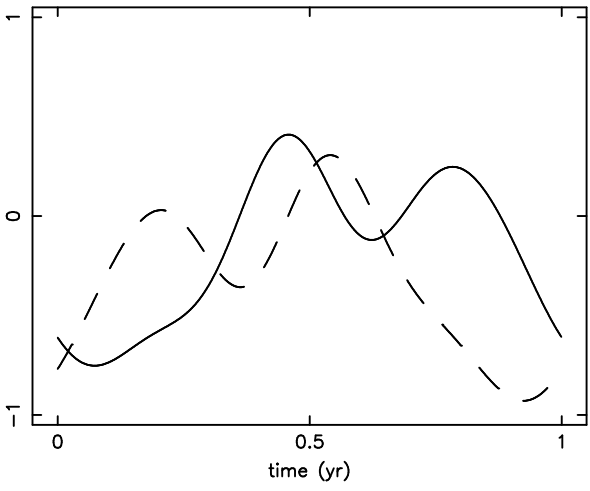
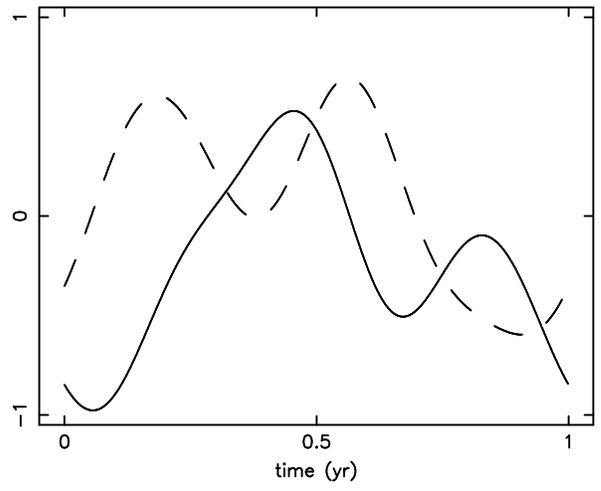
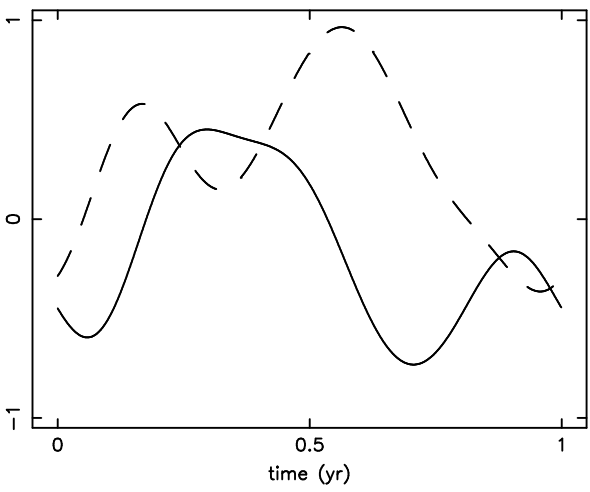
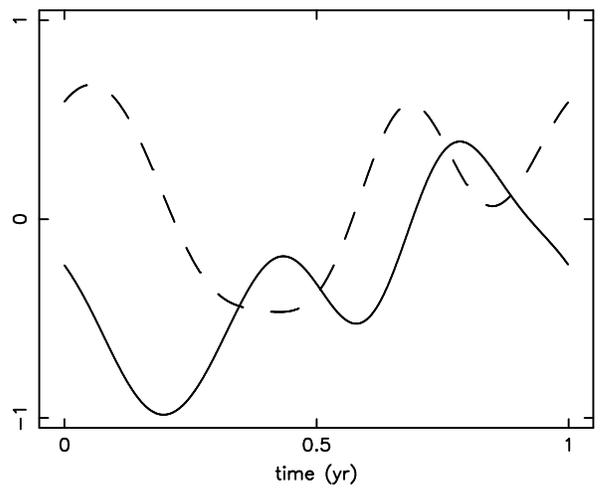
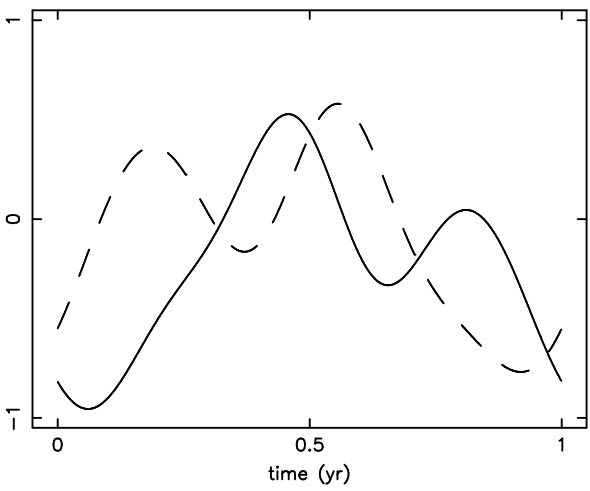
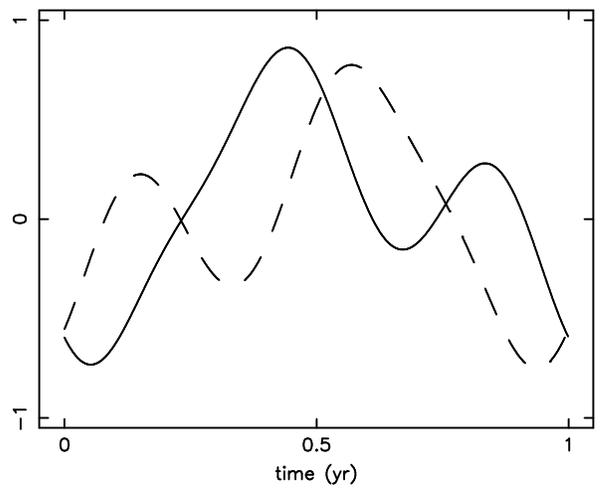
\end{document}